\documentclass[physics]{ejs_author} 
\usepackage{upgreek}
\usepackage{amsmath, amsthm, amssymb, amsfonts, mathrsfs}  
\usepackage{latexsym}
\usepackage{mathdots}
\usepackage{amssymb}
\usepackage{graphicx,bm,units,yfonts,yhmath,amscd}

\usepackage{lineno}
\title{The Non-Commutative Geometry of the Complex Classes of Topological Insulators}
\shorttitle{NCG of the Complex Classes of Topological Insulators}

\articletype{Research Article}

\year{2013}
\license{Emil Prodan}
\author{Emil Prodan\email{prodan@yu.edu}}

\institute{
      Department of Physics, Yeshiva University,\\
     245 Lexington Av, 10016 New York, USA\\
       -----\\
     \bigskip \emph{``It is easer to find a needle in a hay stack ... if someone points to it."\\
     \smallskip   Author: unknown.}
          }
\abr{TQM}
\journal{Topological Quantum Matter}

\abstract{Alain Connes' Non-Commutative Geometry program \cite{Connes:1994wk} has been recently carried out \cite{ProdanJPA2013hg,ProdanOddChernArxiv2014} for the entire A- and AIII-symmetry classes of topological insulators, in the regime of strong disorder where the insulating gap is completely filled with dense localized spectrum. This is a short overview of these results, whose goal is to highlight the methods of Non-Commutative Geometry involved in these studies.  The exposition proceeds gradually through the cyclic cohomology, quantized calculus with Fredholm-modules, local formulas for the odd and even Chern characters and index theorems for the odd and even Chern numbers. The characterization of the A- and AIII-symmetry classes in the presence of strong disorder and magnetic fields emerges as a natural application of these tools.}

\keywords{topological insulators \*\ non-commutative geometry \*\ aperiodic crystals \*\ A- and AIII-symmetry classes}

\begin{document}
\firstpage{1}
\maketitle

\section{Introduction} It is often said that the topological insulators (TI) are the equivalent of the Quantum Integer Hall Effect (IQHE), but without the need of an external magnetic field (or any other external field for that matter). It will be informative to recall some of the outstanding characteristics of the IQHE \cite{PrangeBook1987cu} and see how they manifest or would manifest in TIs. First, one should recall that disorder is paramount for observing IQHE, since without disorder the widths of the Hall plateaus would be reduced to zero.  Indeed, if the gaps between the Landau bands were clean, then the Fermi level $E_F$ will jump from one Landau band to another when increasing or decreasing the electron density and never have a change to be inside the gaps where the quantization occur.  The ideal picture of a topological insulator, that most of us have, is that of a pristine material with a clean insulating gap and with $E_F$ fixed right in the middle of this gap. The reality, however, is that the Fermi level never stays in a clean gap but rather pins itself to impurity states which inherently occur in real samples. Instead of leaving the fate of the Fermi level in the hands of such random events, one can stabilize $E_F$ deep in the insulating gap by filling the gap with dense localized spectrum (Fermi level engineering of TIs via dopants works this way). And if it is to ever observe quantized plateaus in the magneto-electric response of a strong TI as one varies the electron density via a gate potential, then $E_F$ must again be embedded in dense localized spectrum, otherwise $E_F$ will simply jump into the conduction or the valence spectrum upon gating. The inescapable conclusion is that, like for IQHE, the understanding of the regime where $E_F$ is not in a clean gap but is embedded in dense localized spectrum is central to the physics of topological materials. As such, the stability of topological phases under strong disorder should be placed among the key issues in a complete theory of TIs. 

Another hallmark of IQHE is the spike of the direct conductivity whenever the system crosses from one quantized plateau to another, which indicates that $E_F$ crosses a region of delocalized energy spectrum. Since this occurs in a 2-dimensional system at strong disorder, one cannot help but contemplate what an interesting effect this is! Viewed from a particular angle, however, this effect is not that un-expected, because what it shows is that the different topological IQHE phases, characterized by the integer values of Hall conductance, are truly separated by a quantum transition, the Anderson localization-delocalization transition. If  the topological phases described in the classification table of topological insulators and superconductor (see Table 3 in Ref.~\cite{RyuNJP2010tq}) are indeed macroscopically discernible, such quantum transitions must also occur at the crossings between these phases. 

While the above characteristics are extremely well understood for IQHE, both theoretically and experimentally, they are poorly understood for TIs. For example, the quantum critical regime of a topological-to-trivial insulator transition has never been achieved experimentally, for any class whatsoever. On the rigorous theoretical front, most of the periodic table of TIs remains un-explored under the conditions of strong disorder. For example, we do not have a formula for the $\bm Z_2$ invariant of the strong topological insulators in 3-dimensions when the Fermi level is embedded in dense localized spectrum, and in fact the fate of the $\bm Z_2$ invariant is not known in those conditions. The very few numerical experiments \cite{FulgaPRB2012cg,LeungPRB2012vb,SbierskiArxiv2014vh}, based on plain extensions of the translationally invariant formulas of the $\bm Z_2$ invariant, were conducted for extremely small system sizes and are far from being conclusive. This brings us to the main question we want to address: Is it possible to push the classification of topological insulators and superconductors, as it appears in the table put forward by Refs.~\cite{SchnyderPRB2008qy,kitaev:22,RyuNJP2010tq}, into the regime of strong disorder? Does the table stays the same or it needs to be modified?

Now, an important observation is the following simple but fundamental principle: If the bulk topological invariant labeling the different phases in the table stays quantized and non-fluctuating as long as the Anderson localization length is finite, then the characteristics discussed above are necessarily present. Indeed, this property will ensure, on one hand, the stability of the topological phases in the presence of strong disorder and, on the other hand, that the only way to cross from one topological phase to another is via a divergence of the localization length. For IQHE,  the Hall conductance was proven to posses this property using non-commutative geometry in the 1990's \cite{BELLISSARD:1994xj}. This result represents one of the most important applications of the non-commutative geometry in condensed matter physics (cf \cite{Connes:1994wk} pg. 363). It was only recently that similar mathematically rigorous results start appearing for other topological phases \cite{ProdanJPA2013hg,ProdanOddChernArxiv2014}. They gave a complete characterization, from the bulk perspective, of the entire complex classes (the A- and AIII-symmetry classes in any dimension) of the classification table \cite{RyuNJP2010tq} of topological insulators and superconductors. As such, the methods of non-commutative geometry have been extended from the upper-left corner of this table to the entire rows of the complex classes, leading to an affirmative answer to the main question posed above. 

When discussing these results with his colleagues, the author is often asked how crucial is Alain Connes' non-commutative geometry for the whole development? Of course, after understanding the arguments and seeing the final conclusions, one can reproduce them via different methods. However, without the guidance from non-commutative geometry, searching for the correct form of the index theorems would have been like searching for a needle in a haystack. To convince the reader of this fact, the paper presents first some key elements of the non-commutative geometry program, which as we shall see lay down the basic principles and the guiding philosophy. Then the paper gradually builds the specific structures needed for the problem at hand. The quantization and the homotopy stability of the topological invariants for the complex classes of topological insulators, together with the general conditions when these happen, will then naturally emerge.

\section{Elements of $K$-Theory} 

For periodic insulators in dimension $d$, the space of the occupied electron states generates a complex vector-bundle over the Brillouin torus $\mathbb T^d$. It is quite often in condensed matter physics that two independent models are put together to generate a new model, with a number of Bloch bands equal to the sum of those in the original models (all we have to do is to assign different labels to the bands of the two models, such as spin-up and spin-down, or the $s$-character and the $p$-character and so on). This operation defines an addition on the space of vector-bundles. For topological insulators, a model is rather seen as representing all the models which are isomorphic to it (that is, for each $\bm k$ one can define an invertible linear map between the Bloch vectors of the two models, and this linear map is continuous of $\bm k$). The addition operation makes the space of isomorphically equivalent vector bundles over the Brillouin torus into a semigroup. This semigroup can be canonically completed to a group (via the Grothendieck completion) and this group is call the $K_0$ group of the base manifold, which in our case is just the Brillouin torus. Two band-structures belong to the same equivalence class of $K_0(\mathbb T^d)$ group if one can add a number of flat bands to each of the band structures so that to make them isomorphic with each other. This is exactly the topological equivalence which is sought for the classification of the topological insulators. Why classification by the $K$-groups and not by homotopy of homology which are more refined? The lattice models only capture the electronic structure near the Fermi level, which is important for the typical experiments in condensed matter physics. The lattice models can be made more and more precise by increasing the number of molecular orbital states per unit cell (that is, the $N$ in $\ell(\mathbb Z^d,\mathbb C^N$)), but typically the precision of the experiments can be matched using only a small number of molecular states (or number of bands for translational invariant models). While there is clearly a liberty in setting up the models, the topological classification of the physical systems themselves should be independent of the models we use. It is then fairly obvious that we need to classify by the $K$-theory, which is stable against augmentation with trivial bundles (while the classification by homotopy or homology is not). This is perhaps best explained in Ref.~\cite{StoneJPA2011jv}.

Condensed matter physicists are familiar with numerical invariants (such as the first Chern number) which assign numerical values (in $\mathbb Z$ or $\mathbb Z_2$ for example) to the equivalent band structures. But $K_0(\mathbb T^d)$ is a topological invariant itself since, as a group, it cannot be changed by continuous deformations of the base manifold. In fact, the $K$-groups are among the primordial topological invariants of a topological space. The numerical invariants which are concretely evaluated in applications are just morphisms from the $K$-groups to the simple numerical groups such as $\mathbb Z$ or $\mathbb Z_2$. This is pretty much the big picture one must have in mind.

Since the present interest is in the disordered case, one needs a more algebraic description of the $K$-groups, which for classical topological spaces can be found in the excellent monograph by Park  \cite{ParkBook2008fh}. In essence, the band structures can be equivalently described by the projectors onto the occupied electron states. Any complex vector-bundle can be constructed as the range of a projector continuously defined over the base manifold. These projectors are embedded in the space of infinite matrices (zero elements are added for this purpose). Furthermore, two projectors $P$ and $P'$, defined over $\mathbb T^d$, are said to be equivalent if there exists an invertible element $S$, also defined over $\mathbb T^d$, such that $P' = SPS^{-1}$ or equivalently, if there is a continuous homotopy $P_t$ of projectors such that $P_0 = P$ and $P_1=P'$. The equivalence classes are denoted by $[P]$. An addition operation can be defined:
\begin{equation}
[P]+[Q]=\left [
\begin{array}{cc}
P & 0 \\
0 & Q
\end{array}
\right ],
\end{equation}
and the space of equivalence classes of projectors become a semigroup. Then the $K_0$ group can be equivalently defined as the Grothendieck completion of this semigroup. In even dimensions, the (even-) Chern number \cite{ParkBook2008fh}:
\begin{equation}\label{EvenChernNr1}
\mathrm{Ch}_d (P)=\frac{(\frac{1}{2}(d-1))!}{(2 \pi \imath)^{\frac{d+1}{2}}d !} \int\limits_{\mathbb T^d}\mathrm{Tr}\left \{ \big ( P\bm d P\wedge \bm d P \big )^\frac{d}{2} \right \} \in \mathbb Z,
\end{equation}
which is constant on the whole equivalence class of $P$ in $K_0(\mathbb T^d)$, defines a group morphism from $K_0$ to $\mathbb Z$.

The $K_1$ group is constructed from the invertible matrices defined over the base manifold. More precisely, let $\mathrm{GL}_\infty(\mathbb T^d)$ be the inductive limit (with the natural embeddings) of the groups $\mathrm{GL}_k(\mathbb T^d)$ of $k \times k$ invertible matrices continuously defined over the Brillouin torus. Elements from $\mathrm{GL}_\infty(\mathbb T^d)$ can be thought as invertible infinite matrices which, except for a finite upper-left corner, have only 1 on the diagonal. Then $K_1(\mathbb T^d)$ is defined as $\mathrm{GL}_\infty(\mathbb T^d)/\mathrm{GL}_\infty(\mathbb T^d)_0$ where $\mathrm{GL}_\infty(\mathbb T^d)_0$ is the inductive limit of $\mathrm{GL}_k(\mathbb T^d)_0$, the connected component of the unity in $\mathrm{GL}_k(\mathbb T^d)$. Two invertibles belong to the same class of $K_1(\mathbb T^d)$ if they are homotopic to each other. This group is usually seen as classifying the maps between the vector-bundles constructed over a base manifold. However, the ground state of a condensed matter system from the AIII-symmetry class is uniquely determined by a unitary matrix (the off-diagonal sector of the flat-band Hamiltonian, see \cite{SchnyderPRB2008qy,Schnyder:2009qa,RyuNJP2010tq}). As such, $K_1(\mathbb T^d)$ classifies the AIII-symmetric systems. In odd-dimensions, the classical odd-Chern number \cite{ParkBook2008fh}:
\begin{equation}\label{OddChernNr1}
\widetilde {\mathrm{Ch}}_d (U)=\frac{(\frac{1}{2}(d-1))!}{(2 \pi \imath)^{\frac{d+1}{2}}d !} \int\limits_{\mathbb T^d}\mathrm{Tr}\left \{ \big ( U^{-1} {\bf d} U \big )^d \right \} \in \mathbb Z,
\end{equation}
which is constant on the whole equivalence class of $U$ in $K_1(\mathbb T^d)$, defines a group morphism  from $K_1$ to $\mathbb Z$.

When disorder and magnetic fields are present, the Brillouin torus continue to have a meaning as a non-commutative space $\mathcal A$, more precisely, as a non-commutative $C^*$-algebra endowed with a non-commutative differential calculus (see Chapter~\ref{NCBT}). As one shall see, the task becomes the classification of the projectors and of invertible elements from the non-commutative $C^*$-algebra $\mathcal A$, which are contained in the topological $K_0^\mathrm{T}(\mathcal A)$ and $K_1^\mathrm{T}(\mathcal A)$ groups, respectively. We specifically placed the label ``topological" because these groups can be defined in a purely algebraic fashion, which will play a role in our exposition and will be introduced shortly. A pedagogical introduction to $K$-theory of $C^*$-algebras can be found in the excellent monographs by Wegge-Olsen \cite{WeggeOlsenBook1993de} or by Rordam, Larsen and Laustsen \cite{RordamBook2000bv}. Since these groups play a central role in our story, it is worth spending a few lines explaining their definition. Let us start with $K_0$. A projector in a $C^*$ algebra is an element which obeys $p^2=p$ and $p^* = p$. Let $\mathcal P(\mathcal A)$ denote the set of all projectors in $\mathcal A$. In fact, let $M_{n,m}(\mathcal A)$ be the algebra of $n \times m$ matrices with entries from $\mathcal A$ and let $\mathcal P_n(\mathcal A)$ denote the set of projectors from $M_{n,n}(\mathcal A)$. Consider the infinite union (where the $\mathcal P_n(\mathcal A)$'s are considered pairwise disjoint): 
\begin{equation}
\mathcal P_\infty (\mathcal A) = \cup_{n=1}^\infty \mathcal P_n(\mathcal A)
\end{equation}
together with the equivalence relation:
\begin{equation}\label{Equvalence1}
\mathcal P_n(\mathcal A) \ni p \sim_0 q \in \mathcal P_m(\mathcal A) \ \iff \left \{
\begin{array}{l}
p = vv^* \\
 q=v^*v
 \end{array}
 \right .
\end{equation}
for some matrix $v$ from $M_{m,n}(\mathcal A)$. Note that this equivalence relation is purely algebraic and can be considered for algebras without a norm. For $C^*$-algebras however, the $\sim_0$ and the homotopy equivalence of two projectors is the same. The following addition operation on $\mathcal P_\infty$:
\begin{equation}
p+q = \left (
\begin{array}{cc}
p & 0 \\
0 & q
\end{array}
\right )
\end{equation}
is compatible with the equivalence relation $\sim_0$ and $(\mathcal P_\infty(\mathcal A)/ \sim_0, +)$ becomes a semigroup. Then $K_0^T(\mathcal A)$ is defined as the Grothendieck completion of this semigroup. As we shall see, the regime of strong disorder is quite demanding and it will force on us to work with a sub-algebra $\mathcal A_{\mathrm{loc}}$ of the weak von Neumann extension of $\mathcal A$, and there one has to start with the $\sim_0$ equivalence, alone, in which case one talks about the algebraic $K$-group $K_0^\mathrm{A}(\mathcal A_{\mathrm{loc}})$.

The $K_1^\mathrm{T}(\mathcal A)$ group is  defined as before:
\begin{equation}\label{TopK1}
K_1^{\mathrm{T}}(\mathcal A)=\mathrm{GL}_\infty(\mathcal A)/\mathrm{GL}_\infty(\mathcal A)_0,
\end{equation}
where, very much like before, $\mathrm{GL}_\infty(\mathcal A)$ represents the inductive limit of the groups $\mathrm{GL}_k(\mathcal A)$ of $k \times k$ invertible matrices with entries from $\mathcal A$, with $\mathrm{GL}_k $ naturally embedded in $\mathrm{GL}_{k+1}$ as $\left ( \begin{array}{cc} \mathrm{GL}_k & 0 \\ 0 & 1 \end{array} \right )$. The space $\mathrm{M}_\infty(\mathcal A)$ of square matrices that are not necessarily invertible is constructed the same way. As before, $\mathrm{GL}_\infty(\mathcal A)_0$ denotes the inductive limit of $\mathrm{GL}_k(\mathcal A)_0$, the connected component of the unity in $\mathrm{GL}_k(\mathcal A)$. As we mentioned, we also need to consider the algebraic $K_1$-group, which for a generic $*$-algebra $\mathcal A_{\mathrm{loc}}$ is defined as:
\begin{equation}\label{AlgK1}
K_1^{\mathrm{A}}(\mathcal A_{\mathrm{loc}})=\mathrm{GL}_\infty(\mathcal A_{\mathrm{loc}})\ / \ [\mathrm{GL}_\infty(\mathcal A_{\mathrm{loc}}),\mathrm{GL}_\infty(\mathcal A_{\mathrm{loc}})],
\end{equation}
where $[\mathrm{GL}_\infty(\mathcal A_{\mathrm{loc}}),\mathrm{GL}_\infty(\mathcal A_{\mathrm{loc}})]$ is the normal subgroup of commutators, generate by products of the form $fgf^{-1}g^{-1}$.

Non-commutative geometry's primary target is the $K$-theory of non-commutative spaces. The core of the formalism can and is developed at a purely algebraic level (see the discussion at pg.~180 in Ref.~\cite{Connes:1994wk}), through the algebraic $K$-groups. The topological invariants defined at the pure algebraic level are then extended to the topological $K$-groups (which in many cases may differ from the algebraic ones) using hard functional analysis. 

Below is a summary of the main points of the $K$-theoretic framework:
\begin{itemize} 
\item The $K$-groups are the among the primordial topological invariants of a topological space.
\item The numerical invariants such as the even- and odd-Chern numbers are just morphisms from the $K$-groups to a numerical abelian group.
\item Specifically, the $K_0(\mathbb T^d)$ and $K_1(\mathbb T^d)$ groups are the primordial topological invariants for periodic crystals from the A and AIII-symmetry classes, respectively.
\item The even- and odd-Chern numbers are numerical invariants which define morphisms from $K_{0/1}(\mathbb T^d)$ groups to $\mathbb Z$ for even/odd space-dimensions $d$, respectively. 
\item In the presence of disorder and magnetic fields, the Brillouin torus is replaced by the non-commutative Brillouin torus, defined as a non-commutative $C^*$-algebra $\mathcal A$ endowed with a non-commutative differential calculus. 
\item The classical $K$-groups are replaced by the $K_{0/1}(\mathcal A)$ groups of this algebra, which classify the ground states of the disordered systems from the A and AIII-symmetry classes, respectively. 
\item Non-commutative geometry is used to generate morphisms from the $K$-groups to abelian numerical groups (such as $\mathbb Z$). These morphisms are nothing but the numerical invariants which can be computed explicitly.
\item The invariants are typically defined on sub-algebras of $\mathcal A$ and for the algebraic $K$-groups. Extending these invariants to the topological $K$-groups requires additional work.
\end{itemize}

\section{Elements from Alain Connes' Non-Commutative Geometry Program}

It is important to project the generality of the non-commutative geometry, hence the theory will be presented in the most general and abstract setting possible, by closely following the presentation in Alain Connes' monograph \cite{Connes:1994wk}. As such, the discussion will be about a generic $*$-algebra, denoted by $\mathcal C$, about its $K_{0,1}^\mathrm{A}(\mathcal C)$ groups and the morphisms from these $K$-groups to a numerical abelian group. According to the previous Chapter, these are the sought numerical invariants.

\subsection{Cyclic Cohomology}

The cyclic cohomology was introduced first in Refs.~ \cite{ConnesIHES1982hf,ConnesIHES1983tr}. For a generic $*$-algebra $\mathcal C$, it is defined as the cohomology of the complex $(C_\lambda^n(\mathcal C),b)$:
\begin{equation}
\ldots \stackrel{b}{\rightarrow} C_\lambda^{n-1}(\mathcal C) \stackrel{b}{\rightarrow} C_\lambda^n(\mathcal C) \stackrel{b}{\rightarrow} \ldots
\end{equation}
 where $C_\lambda^n(\mathcal C)$ is the space of cyclic $(n+1)$-linear functionals on $\mathcal C$: 
\begin{equation}
\phi(c_1,c_2,\ldots,c_n,c_0)=(-1)^n \phi(c_0,c_1,\ldots,c_n),
\end{equation}
and $b:C_\lambda^n(\mathcal C) \rightarrow C_\lambda^{n+1}(\mathcal C)$ is the Hochschild coboundary map:
\begin{align}
b \phi(c_0,c_1,\ldots,c_{n+1}) = & \sum_{j=0}^n (-1)^j \phi(c_0,\ldots, c_j c_{j+1}, \ldots c_{n+1}) \bigskip \nonumber \\
& +(-1)^{n+1} \phi(c_{n+1}c_0, \ldots, c_n). 
\end{align}
An element $\varphi$ from $C_\lambda^n(\mathcal C)$ is said to be an $n$-cyclic cocycle if it satisfies $b\varphi =0$. Such elements play the role of closed differential forms in the classical de Rham cohomology. The cohomology class of $\varphi$ in the complex $(C_\lambda^n(\mathcal C),b)$, which contains all $\varphi'$ with $\varphi - \varphi' = b \phi$, will be denoted by $[\varphi]$. A cyclic cocycle is odd/even if $n$ is an odd/even integer. The interest in the cyclic cocycles, at least for the present context, comes from the fact that they pair well with $K_{0,1}^\mathrm{A}(\mathcal C)$ groups (cf. \cite{Connes:1994wk}, pg. 224).

\begin{proposition}[Pairing odd cyclic cocycles with $K_1$]\label{CyclicCocycles1}
Let $\varphi$ be an odd cyclic cocycle, and let $\varphi \# \mathrm {Tr}$ be its natural extension to $\mathrm{M}_\infty(\mathcal C)$. Then the map 
\begin{equation}\label{Cocycle1}
\mathrm{GL}_\infty(\mathcal C) \ni v \rightarrow (\varphi \# \mathrm {Tr}) (v^{-1}-1,v-1,\ldots,v^{-1}-1,v-1)
\end{equation}
is constant on the equivalence class $[v]$ of $v$ in $K_1^\mathrm{A}(\mathcal C)$. Furthermore, $\varphi$ can be replaced by any other representative from its cohomology class. As such, there exists a natural pairing between  $K_1^\mathrm{A} (\mathcal C)$ and the odd cohomology of $\mathcal C$:
\begin{equation}\label{Pairing1}
\langle [v],[\varphi] \rangle =(\varphi \# \mathrm{Tr}) (v^{-1}-1,v-1,\ldots,v^{-1}-1,v-1).
\end{equation}
The pairing is not necessarily integral.
\end{proposition} 

\begin{proposition}[Pairing even cyclic cocycles with $K_0$]\label{CyclicCocycles2}
Let $\varphi$ be an even cyclic cocycle, and let $\varphi \# \mathrm {Tr}$ be its natural extension to $\mathrm{M}_\infty(\mathcal C)$. Then the map 
\begin{equation}\label{Cocycle2}
\mathcal P_\infty(\mathcal C) \ni p \rightarrow (\varphi \# \mathrm {Tr}) (p,p,\ldots,p)
\end{equation}
is constant on the equivalence class $[p]$ of $p$ in $K_0^\mathrm{A}(\mathcal C)$. Furthermore, $\varphi$ can be replaced by any other representative from its cohomology class. As such, there exists a natural pairing between  $K_0^\mathrm{A} (\mathcal C)$ and the even cohomology of $\mathcal C$:
\begin{equation}\label{Pairing2}
\langle [p],[\varphi] \rangle =(\varphi \# \mathrm{Tr}) (p,p,\ldots,p).
\end{equation}
The pairing is not necessarily integral.
\end{proposition} 

While the above statements are formulated for the algebraic $K$-theory, an immediate Corollary about the homotopy invariance of the cyclic cocycles can be established. Indeed, according to Ref.~\cite{Connes:1994wk} pg.~226, if $\mathcal C$ is a locally convex topological algebra, and if the cyclic cocycle is {\emph continuous} with respect its topology, then the maps defined in Eq.~\ref{Cocycle1} and \ref{Cocycle2} are constant on the homotopies of $v$ and $p$, respectively. In other words, the algebraic $K_{0,1}^\mathrm{A} (\mathcal C)$ groups can be replaced by the topological $K_{0,1}^\mathrm{T} (\mathcal C)$ groups in the Propositions~\ref{CyclicCocycles1} and \ref{CyclicCocycles2}.

\subsection{Quantized Calculus with Fredholm-modules} 

The quantized calculus with Fredholm-modules is of interest for the present context because it is a standard way to generate cyclic cocycles.

\subsubsection{The odd Fredholm-modules} An odd Fredholm-module $(\mathcal{H},F)$ over a $*$-algebra $\mathcal C$ is defined by (cf. \cite{Connes:1994wk} pg.~288) :
\begin{itemize}
\item A representation $\pi$ of $\mathcal C$ in a Hilbert space $\mathcal{H}$;
\item An operator $F$ on $\mathcal H$ with the properties: 
\begin{enumerate}
\item $F^\dagger =F$
\item $F^2=I$ 
\item $[F,\pi(c)]=$ compact for any $c \in \mathcal C$.
\end{enumerate}
\end{itemize}
 If $q$ is a real number larger than or equal to $1$, then a Fredholm-module is said to be $q$-summable over $\mathcal C$ if $[F,\pi(c)]$ belongs to the $q$-th Schatten class for all $c \in \mathcal C$ (that is, the trace of $|[F,\pi(c)]|^q$ is finite).

Now consider an odd integer $n$ and an odd Fredholm-module $(\mathcal{H},F)$ which is $(n+1)$-summable. The quantized calculus of the odd Fredholm-module consists of:
\begin{itemize}
\item The graded algebra $({\bm \Omega},d)$, where ${\bm \Omega} = \bigoplus {\bm \Omega}^k$ with:
$${\bm \Omega}^k = \mathrm{spann}\{ c_0[F,c_1] \ldots [F,c_k], \ c_j \in \mathcal{C}\},$$
\item The differentiation:
$$ {\bm \Omega}^k \ni \eta \rightarrow d \eta = F \eta -(-1)^k \eta F,$$
\item And the closed graded trace:
$${\bm \Omega}^{n}\ni \eta \rightarrow \mathrm{Tr}'\{\eta\}=\nicefrac{1}{2} \ \mathrm{Tr}\{F d\eta\}.$$
\end{itemize}
Then the cyclic $(n+1)$-linear functional:
\begin{align}\label{OddChernCharacter}
\widetilde \tau_{n}(c_0,c_1, & \ldots,c_{n}) = \nonumber \\
& \nicefrac{ (-1)^{n}}{2^{n+1}} \ \mathrm{Tr}'(\pi(c_0) [F,\pi(c_1)], \ldots, [F,\pi(c_{n})]\}
\end{align}
is well defined due to the summability condition and represents an odd cyclic cocycle. Its cohomology class is called the odd Chern character of the Fredholm-module, and is denoted by $\widetilde{\mathrm{Ch}}_*(\mathcal H,F)$ (the tilde is used to distinguish the odd from the even Chern character, defined below). Furthermore, $\widetilde{\mathrm{Ch}}_*(\mathcal H,F)$ pairs well with the algebraic $K_1^\mathrm{A}(\mathcal{C})$ group (see Eq.~\ref{Pairing1}) and the paring is integral (cf. \cite{Connes:1994wk} pg. 296):
\begin{equation}\label{OddIntegrality}
\langle [v],\widetilde{\mathrm{Ch}}_*(\mathcal H,F) \rangle = \mathrm{Index} \ E \pi(v) E \ \in \mathbb{Z},
\end{equation}
where $E$ is the idempotent $E=\frac{1}{2}(1+F)$, as naturally extended over $\mathrm{M}_\infty (\mathcal C) = \mathcal C \otimes \mathrm{M}_\infty(\mathbb{C})$ via $E\otimes 1$. The representation $\pi$ is also extended in a similar way. We recall that, if $\mathcal C$ is a locally convex topological algebra and $\widetilde \tau_{n}$ is continuous in its topology, then $K_1^\mathrm{A}(\mathcal C)$ can be replaced by $K_1^\mathrm{T}(\mathcal C)$.

\subsubsection{The even Fredholm-modules} An even Fredholm-module $(\mathcal{H},F,\gamma)$ over a $*$-algebra $\mathcal C$ is defined by (cf.~\cite{Connes:1994wk} pg.~288):
\begin{itemize}
\item A representation $\pi$ of $\mathcal C$ in a Hilbert space $\mathcal{H}$;
\item An operator $F$ on $\mathcal H$ with the properties:
\begin{enumerate}
\item $F^\dagger =F$
\item $F^2=I$ 
\item $[F,\pi(c)]=$ compact for any $c \in \mathcal C$;
\end{enumerate}
\item A grading $\gamma$ ($\gamma^\dagger = \gamma$, $\gamma^2=1$) such that:
\begin{enumerate}
\item $\gamma \pi(c)= \pi(c) \gamma$ for all $c \in \mathcal C$
\item $\gamma F = -F \gamma.$
\end{enumerate}
\end{itemize}
Again, if $q$ is a real number larger than or equal to $1$, then a Fredholm-module is said to be $q$-summable if $[F,\pi(c)]$ belongs to the $q$-th Schatten class for all $c \in \mathcal C$.

Now consider an even integer $n$ and an even Fredholm-module $(\mathcal{H},F)$ which is $(n+1)$-summable. The quantized calculus of the Fredholm-module consists of:
\begin{itemize}
\item The graded algebra $({\bm \Omega},d)$, where ${\bm \Omega} = \bigoplus {\bm \Omega}^k$ with:
$${\bm \Omega}^k = \mathrm{spann}\{ c_0[F,c_1] \ldots [F,c_k], \ c_j \in \mathcal{C}\},$$
\item The differentiation:
$$ {\bm \Omega}^k \ni \eta \rightarrow d \eta = F \eta -(-1)^k \eta F,$$
\item And the closed graded trace:
$${\bm \Omega}^{n}\ni \eta \rightarrow \mathrm{Tr}'\{\eta\}=\nicefrac{1}{2} \ \mathrm{Tr}\{\gamma F d\eta\}.$$
\end{itemize}
Then the cyclic $(n+1)$-linear functional:
\begin{align}\label{EvenChernCharacter}
\tau_{n}(c_0,c_1, & \ldots,c_{n}) = \nonumber \\
& \nicefrac{ (-1)^{n}}{2^{n+1}} \ \mathrm{Tr}'(\pi(c_0) [F,\pi(c_1)], \ldots, [F,\pi(c_{n})]\}
\end{align}
is well defined due to the summability condition and represents an even cyclic cocycle. Its cohomology class is called the even Chern character of the Fredholm-module, and is denoted by $\mathrm{Ch}_*(\mathcal H,F)$. Furthermore, $\mathrm{Ch}_*(\mathcal H,F)$ pairs well with the algebraic $K_0^\mathrm{A}(\mathcal{C})$ group (see Eq.~\ref{Pairing2}) and the paring is integral (cf. \cite{Connes:1994wk} pg. 296):
\begin{equation}\label{EvenIntegrality}
\langle [p],\mathrm{Ch}_*(\mathcal H,F) \rangle = \mathrm{Index} \ \pi^-(p) F \pi^+(p) \ \in \mathbb{Z},
\end{equation}
where $F$ is naturally extended over $\mathrm{M}_\infty (\mathcal C) = \mathcal C \otimes \mathrm{M}_\infty(\mathbb{C})$ via $F\otimes 1$ and $\pi^\pm$ is the decomposition with respect to the grading $\gamma$ of the similarly extended representation $\pi$. We recall that, if $\mathcal C$ is a locally convex topological algebra and $\tau_{n}$ is continuous in its topology, then $K_0^\mathrm{A}(\mathcal C)$ can be replaced by $K_0^\mathrm{T}(\mathcal C)$.

\subsection{Discussion}

One can generate topological invariants by playing with the algebra and the Fredholm-modules over it, and the possibilities are really endless. To apply the techniques to a specific problem, one needs to define an appropriate algebra and an appropriate Fredholm-module over this algebra. For the problem of classification of complex classes of disordered topological insulators, where the cyclic cocycles have to reproduce the classical Chern and winding numbers, this was accomplished in Ref.~\cite{ProdanJPA2013hg} and \cite{ProdanOddChernArxiv2014}. As we shall see, the constructions are quite natural.

We left out an important chapter of the non-commutative geometry program, namely, obtaining local formulas for the Chern characters (cf.~\cite{ConnesGFA1995re}). We did this on purpose because, as we shall see, in the present context these local formulas emerge from two explicit geometrical identities, which make the whole process extremely transparent. To understand the importance of this aspect, note that the definition of the Chern characters are highly non-local, which is a huge inconvenience. For periodic crystals, for example, these formulas will involve convolutions over the Brillouin torus of highly non-local kernels. In contradistinction, the classical Chern number formulas in Eqs.~\ref{EvenChernNr1} and \ref{OddChernNr1} involve only derivatives and plain products, which are local in character. Showing how to obtain local formulas from the non-local expressions of the Chern characters were among the key results of Refs.~\cite{ProdanJPA2013hg,ProdanOddChernArxiv2014}.

For convenience, we summarize below the important points (for the present context) of the theory presented in this Chapter:
\begin{itemize}
\item Cyclic cocycles pair with the $K$-groups and generate numerical invariants over the equivalence classes from the $K$-groups. These invariants are not necessarily integers. 
\item The quantized calculus with a Fredholm-module produces the Chern character, which is a cyclic cocycle pairing integrally with the $K$-groups. In other words, the numerical invariants generated by the Chern character over the equivalence classes from the $K$-groups do take integer values.
\item For a specific application, one needs to define the appropriate algebra and the appropriate Fredholm-module.
\item The pairing between the Chern characters and the $K$-groups is highly non-local. Obtaining local formulas for the Chern characters is an important step of the non-commutative geometry program. 
\end{itemize}

The remaining Chapters describe how the above machinery is applied to the classification of the topological phases from the A- and AIII-symmetry classes of topological insulators.

\section{The Non-Commutative Brillouin Torus}\label{NCBT}

In this Chapter we define the appropriate algebra and show how this algebra is transformed into a non-commutative manifold via a non-commutative differential calculus. The resulting non-commutative space is called the non-commutative Brillouin torus. The notion of non-commutative Brillouin torus (not to be confused with the non-commutative torus) was introduced in the inspiring work of Jean Bellissard \cite{BellissardLNP1986jf}, who developed an entire non-commutative geometry program for aperiodic solids \cite{BellissardLN2003bv}. The non-commutative Brillouin torus is a true gift to the condensed matter physics, since it enables one to define an equivalent Bloch-Floquet calculus for homogeneous aperiodic systems. So far, every important formula written in $k$-space for periodic crystals, has been ported and evaluated on the non-commutative Brillouin torus for disordered crystals under magnetic fields. Some examples are: Kubo-formula \cite{BELLISSARD:1994xj,Schulz-Baldes:1998vm,Schulz-Baldes:1998oq}, electric polarization formula \cite{Schulz-BaldesCMP2013gh,DeNittisJPA2013fd}, orbital magnetization formula \cite{Schulz-BaldesCMP2013gh}, the even and odd Chern numbers \cite{BELLISSARD:1994xj,ProdanJPA2013hg,ProdanOddChernArxiv2014}, the spin-Chern numbers \cite{Prodan:2009oh}, the magneto-electric response tensor \cite{LeungJPA2012er}, and the winding numbers \cite{MondragonShemArxiv2013ew}. 

\subsection{Homogeneous aperiodic lattice models}
 
Let us now discuss the generic lattice models which describe electrons' dynamics in homogeneous aperiodic materials. The Hilbert space is the set of square-summable functions with $N$ components (= number of molecular orbitals per unit cell), defined over the lattice $\mathcal L = \mathbb Z^d$. The notation for this space is $\ell^2(\mathcal L,\mathbb C^N)= \ell^2(\mathcal L)\otimes \mathbb C^N$. The natural basis for this space is denoted by $\{e_{\bm x}^\alpha\}_{\bm x \in \mathcal L}^{\alpha = 1,\ldots N}$. The aperiodicity can be introduced by a magnetic field and/or by random displacements of the atoms. The generic aperiodic lattice Hamiltonians take the form: 
\begin{equation}\label{LatticeHamOmega}
(H_\omega {\bm \psi})({\bm x})=\sum_{{\bm y}\in \mathcal L} e^{\imath {\bm x}\wedge {\bm y}} \ \hat t_{{\bm x},{\bm y}}(\omega) {\bm \psi}({\bm y}),
\end{equation}
where $\hat t_{\bm x,\bm y}(\omega)$ are $N \times N$ matrices with complex entries, $\wedge$ is an anti-symmetric bilinear form incorporating the effect of the magnetic field (whose exact form is not important), and $\omega$ is a random variable from a probability space $\big (\Omega, d \bm P (\omega) \big )$. Each $\omega$ describes a disorder configuration (see below). The system is said to be homogeneous if:
\begin{equation}
\hat{t}_{{\bm x}-{\bm a},{\bm y}-{\bm a}} (\omega)= \hat{t}_{{\bm x},{\bm y}}(\mathfrak{t}_{\bm a} \omega),
\end{equation}
where $\{\mathfrak{t}_{\bm a}\}_{{\bm a} \in \mathcal L}$ are probability-preserving, ergodic automorphisms on $\Omega$, implementing the lattice-translations group. In these conditions, the collection $\{H_\omega\}_{\omega \in \Omega}$ defines a covariant family of operators, in the sense that:
\begin{equation}
U_{\bm a}H_\omega U_{\bm a}^{-1}=H_{\mathfrak{t}_{\bm a}\omega},
\end{equation}
for any magnetic translation:
\begin{equation}
U_{\bm a}{\bm \psi}({\bm x}) = e^{ -\imath {\bm a} \wedge {\bm x}}{\bm \psi}({\bm x}+{\bm a}).
\end{equation}
Then the quadruple $$\big (\Omega, d\bm P, \{\mathfrak{t}_{\bm a}\}_{{\bm a} \in \mathcal L},\{H_\omega\}_{\omega \in \Omega} \big )$$ is said to define a homogeneous aperiodic lattice system \cite{BellissardLNP1986jf} (where it is understood that $\Omega$ is not just a point). These are the lattice systems for which the non-commutative Brillouin torus can be defined.

A concrete example of such system is shown below:
\begin{equation}\label{LatticeHOmegaExample}
\big ( H_\omega{\bm \psi} \big)({\bm x}) =\sum_{{\bm y}\in \mathcal L} (1+\lambda \omega_{x,y})e^{\imath {\bm x}\wedge {\bm y}} \ \hat{t}_{{\bm x}-{\bm y}}{\bm \psi}({\bm y}),
\end{equation}
where $\omega_{\bm x, \bm y}=\omega_{\bm y,\bm x}$ are independent random variables, uniformly distributed in the interval $[-\frac{1}{2},\frac{1}{2}]$, and $\lambda$ defines the disorder strength. The collection of all $\omega=\{ \omega_{{\bm x},{\bm y}}\}$ can be seen as a point in an infinite dimensional configuration space $\Omega$, which can be equipped with the probability measure:
\begin{equation}
d \bm P (\omega)=\prod_{{\bm x},{\bm y}\in \mathcal L} d\omega_{{\bm x},{\bm y}}.
\end{equation}
The natural action of the discrete $\mathbb{Z}^d$ additive group on $\Omega$:
\begin{equation}
(\mathfrak{t}_{\bm a} \omega)_{{\bm x},{\bm y}}=\omega_{{\bm x}-{\bm a},{\bm y}-{\bm a}}, \ a\in \mathcal L,
\end{equation}
acts ergodically and leaves $d \bm P$ invariant.  Lastly, one can directly verify that indeed $\{H_\omega\}_{\omega \in \Omega}$ defines a covariant family of Hamiltonians under the magnetic translations.

The construction of the non-commutative Brillouin torus for the homogenous models discussed above is standard and is described next \cite{BellissardLN2003bv}.

\subsection{The algebra of covariant physical observables}\label{AlgebraA}

 Consider the set $\mathcal{A}_0$ of continuous functions with compact support:
\begin{equation}
f:\Omega \times \mathcal L \rightarrow  M_{N \times N},
\end{equation}
where $M_{N \times N}$ is the space of $N \times N$ complex matrices, and endow it with the algebraic operations:
\begin{equation}\label{AlgRules}
\begin{array}{l}
(f+g)(\omega,{\bm x})=f(\omega,{\bm x})+g(\omega,{\bm x}), \bigskip \\
(f*g)(\omega,{\bm x})=\sum\limits_{{\bm y} \in \mathcal L} e^{\imath {\bm y} \wedge {\bm x}}f(\omega, {\bm y})g(\mathfrak{t}_{{\bm y}}^{-1}\omega,{\bm x}-{\bm y}).
\end{array}
\end{equation} 
Then $\mathcal A_0$ becomes an algebra with a unit, given by:
\begin{equation}
1(\omega, {\bm x})= I_{N \times N} \delta_{{\bm x},{\bm 0}}.
\end{equation}
Each element from $\mathcal{A}_0$ defines a family of covariant, bounded and finite hopping-range operators on $\ell^2(\mathcal L,\mathbb{C}^{N})$, through the fiber-wise representations:
\begin{equation}\label{OpRep}
(\pi_\omega f){\bm \psi}({\bm x}) =\sum_{{\bm y} \in \mathcal L} e^{ \imath {\bm y} \wedge {\bm x}} f(\mathfrak{t}^{-1}_{\bm y}\omega, {\bm x}-{\bm y}){\bm \psi}({\bm y}).
\end{equation}
For example, the Hamiltonian of Eq.~\ref{LatticeHamOmega} is generated by the element:
\begin{equation}
h(\omega,{\bm x})=\hat{t}_{\bm 0,\bm x} (\omega), \ \pi_\omega(h) = H_\omega.
\end{equation}
The reciprocal is also true, that any finite hopping-range family of covariant operators $F_\omega$ on $\ell^2(\mathcal L,\mathbb{C}^{N})$ defines an element from the algebra $\mathcal A_0$, via:
$$f(\omega,{\bm x})_{\alpha \beta} = (e_{\bm 0}^\alpha, F_\omega e_{\bm x}^\beta).$$

The following equation:
\begin{equation}\label{Norm}
||f|| = \sup_{\omega \in \Omega} \ \sup_{||\psi||=1}\sqrt {\langle \pi_\omega(f) \psi, \pi_\omega(f) \psi \rangle }
\end{equation}
defines a  norm on $\mathcal{A}_0$ and: 
\begin{equation}
f^\ast(\omega,{\bm x})=f(\mathfrak{t}_{\bm x}^{-1}\omega,-{\bm x})^\dagger
\end{equation}
defines a $\ast$-operation. Then the completion of $\mathcal{A}_0$ under the norm defined in Eq.~\ref{Norm} becomes a $C^\ast$-algebra, which is denoted by $\mathcal{A}$.

\subsection{The non-commutative differential calculus}

The non-commutative differential calculus over $\mathcal{A}$ is defined by:
\begin{enumerate}
\item Integration:
\begin{equation}\label{Trace}
\mathcal{T}\{f\}=\int_\Omega d \bm P (\omega) \ \mathrm{tr}_\alpha \{f(\omega,{\bm 0})\}, 
\end{equation}
where $\mathrm{tr}_\alpha$ denotes the trace over the orbital degrees of freedom $\alpha$.
\item Derivations ($i=1,\ldots,d$):
\begin{equation}
(\partial_i f)(\omega,{\bm x}) = \imath x^i f(\omega,{\bm x}).
\end{equation}
\end{enumerate}
The triplet $(\mathcal{A},\mathcal{T},\partial)$ defines the non-commutative Brillouin torus. 

Below are a few specific rules of calculus that will be used in the analysis to follow:
\begin{itemize}
\item The integration is cyclic: 
\begin{equation}
\mathcal T\{fg\}=\mathcal T\{gf\}.
\end{equation}
\item The integration by parts holds:
\begin{equation}\label{PartialInt}
\mathcal T\{f \partial_i g\}=-\mathcal T \{(\partial_i f )g\},
\end{equation}
whenever the two integrals are finite.
\item There is the following equivalent formula of calculus:
\begin{equation}\label{Rule1}
{\mathcal T}\{f \ldots g\}=\int_\Omega d\bm P (\omega) \mathrm{tr}_\alpha\big \{  (\pi_\omega f ) \ldots  (\pi_\omega g) \big \}.
\end{equation}
\item The operator representation of the non-commutative derivation is:
\begin{equation}\label{Rule2}
\pi_\omega(\partial_i f)=\imath[X_i,\pi_\omega f].
\end{equation}
\end{itemize} 

\subsection{The sub-algebra of localized observables}

In the regime of strong disorder, when the spectral gaps of the Hamiltonians are filled with localized spectrum, the projectors and the invertible elements characterizing the ground states of the systems from A- and AIII-symmetry classes no longer belong to $\mathcal A$. Instead they belong to the weak von-Neumann closure of $\mathcal A$ denote here by $\mathcal A'$, which for the present context can be described as the closure of $\mathcal A_0$ under the norm:
\begin{equation}||f||' = \bm P-\operatorname*{\mathrm{ess}~\sup}\limits_{\omega \in \Omega} \ \sup_{||\psi||=1}\sqrt {\langle \pi_\omega(f) \psi, \pi_\omega(f) \psi \rangle}.
\end{equation}
If the Fermi level resides in a region of localized spectrum, the kernel of the projectors and of the invertible elements characterizing the ground states of the systems from A- and AIII-symmetry classes decay exponentially when averaged over the disorder. As such, it is natural to define a sub-algebra of what we call the localized observables. This set, together with a proper norm will become the natural domain of definition for the cyclic cocycles. 

\begin{proposition}\label{TildeA'} Consider the set $\mathcal A_{\mathrm{loc}}$ of elements $f $ in the weak von-Neumann closure of $\mathcal A$, obeying the following condition:
\begin{equation}\label{Condition}
\int_\Omega d\bm P (\omega) \  |f(\omega, {\bm x})| \leq A e^{-\lambda |{\bm x}|}, \ \mbox{for some} \ A,\lambda>0.
\end{equation}
Endow this set with the topology induced by the so called GNS norm:
\begin{equation}
||f||_{GNS} = \sqrt{ \mathcal{T}\{ff^*\} } = \sqrt{ \mathcal{T}\{|f|^2 \} }.
\end{equation}
Then:
\begin{enumerate} 
\item The integrals
\begin{equation}
\mathcal T \{ \bm \partial ^{\bm a_1}f_1 \ldots \bm \partial ^{\bm a_k}f_k\}
\end{equation}
are always finite for $f_i$-s from $\mathcal A_{\mathrm{loc}}$.
\item The set $\mathcal A_{\mathrm{loc}}$ is a dense topological $*$-sub-algebra of the weak von-Neumann closure of $\mathcal A$.
\item Any functional of the type (and the natural generalizations):
\begin{equation}
\mathcal A_{\mathrm{loc}} \ni f \rightarrow T(f)=\mathcal T\{(\bm \partial^{\bm a}f)g\}, \ (g \in \mathcal A_{\mathrm{loc}})
\end{equation}
is continuous.
\end{enumerate}
\end{proposition}
\proof All points were proved in \cite{ProdanOddChernArxiv2014}. However, it is interesting to take a closer look at point (3). The continuity of $T$ follows by observing that any such functional can be rewritten as $(-1)^{|\bm a|}\mathcal T\{f(\bm \partial^{\bm a}g)\}$ by using the partial integration (see Eq.~\ref{PartialInt}). Then
\begin{align}\label{C1}
|T(f)-T(f')| & = |\mathcal T\{(f-f')(\bm \partial^{\bm a}g)\}| \\
& \leq \sqrt {\mathcal T\{|\bm \partial^{\bm a}g|^2\} } \sqrt{ \mathcal T\{|f-f'|^2\} }, \nonumber
\end{align}
where Schwartz inequality 
\begin{equation}
|\mathcal T\{fg\}|\leq \sqrt{ \mathcal T\{|f|^2\} \mathcal T\{|g|^2\} }
\end{equation}
 has been used. In the last line of Eq.~\ref{C1} one can immediately identify the GNS norm, to conclude: $|T(f)-T(f')| \leq const. ||f-f'||_{GNS}$. \qed
 
 The continuity in the above proof emerges quite natural, so one can really say that the GNS-norm is indeed the natural norm to consider on $\mathcal A_{\mathrm{loc}}$.
 
\section{Natural Fredholm-modules over the non-commutative Brillouin torus}

\subsection{Clifford algebras and their irreducible representations}

It is really instructive to look at the irreducible representations of the Clifford algebra $C_{n,0}$ defined by:
\begin{equation}
\Gamma_i \Gamma_j + \Gamma_j \Gamma_i = 2\delta_{ij}, \ i,j=1,\ldots,n.
\end{equation}
The irreducible representations of $C_{n,0}$ can be generate by the following procedure. First, one should make a clear distinction between the $n=$ odd and $n=$ even cases. Here, the symbol $\sigma$ will be used for the odd case and the symbol $\gamma$ will be used for the even case, to denote the generators of the irreducible representations. For $n=1$, set $\sigma_1 =1$. Suppose one knows already the generators $\sigma_1$, \ldots, $\sigma_n$ for some odd $n$. Then the irreducible representation of the even $C_{n+1,0}$ is obtained as:
\begin{equation}
\gamma_i = \left (
\begin{array}{cc}
0 & \sigma_i \\
\sigma_i & 0
\end{array}
\right ), \ i=1,\ldots, n,
\end{equation}
and
\begin{equation}
\gamma_{n+1}=\imath \left (
\begin{array}{cc}
0 & - I \\
I  &   0
\end{array}
\right ).
\end{equation}
With the even irreducible representation at hand, one constructs the next odd irreducible representation as:
\begin{equation}
\sigma_i = \gamma_i, \ i=1,n+1,
\end{equation}
and
\begin{equation}
\sigma_{n+2}=(-\imath)^\frac{n+1}{2} \gamma_1 \gamma_2 \ldots \gamma_{n+1}.
\end{equation}
If one starts from $n=1$, he can slowly build all the irreducible representations of $C_{n,0}$ by following the above concrete steps.

Here is what one learns from this exercise. The dimensions of the Hilbert spaces for the irreducible representations of $C_{2m,0}$ and $C_{2m+1,0}$  are both equal to $2^m$. The even representations have a natural grading. Indeed, if one takes $\gamma_0 = \sigma_{2m+1}$, then $\gamma_0^2 = 1$ and $\gamma_0 \gamma_i = - \gamma_i \gamma_0$ for all $i=1,\ldots,2m$. Such grading doesn't exist for the odd Clifford algebras. As such, one can already anticipate that the even/odd Clifford algebras will be used to define even/odd Fredholm-modules, respectively. 

\subsection{The natural Fredholm-modules in odd dimensions}

This chapter describes the natural Fredholm-module over the non-commutative Brillouin torus in odd dimensions $d$ \cite{ProdanOddChernArxiv2014}. Let $\mathrm{Cliff}(d)$ denote the $2^{\frac{d-1}{2}}$-dimensional Hilbert space for the irreducible representation of the odd $C_{d,0}$. The Hilbert space $\mathcal{H}$ of the Fredholm-modules is defined as:
\begin{equation}
{\mathcal H}=\ell^2(\mathcal L,\mathbb{C}^N)\otimes \mathrm{Cliff}(d).
\end{equation}
 The $C^*$-algebra $\mathcal{A}$ can be represented on ${\mathcal H}$ by  the fiber-wise representations $\pi_\omega \otimes 1$, $\omega\in \Omega$. The operator $F$ of the Fredholm-module is defined as the phase of the Dirac operator:
\begin{equation}\label{Dirac}
D=\sum_{i=1}^d X^i \otimes \sigma_i.
\end{equation}
The following shorthands ${\bm v} \cdot {\bm \sigma}=\sum_{i=1}^d v^i \otimes \sigma_i $ and $\hat{{\bm v}}={\bm v}/|{\bm v}|$ will be used throughout. Also,
\begin{equation}\label{TranslatedDirac}
D_{\bm a}=({\bm X}+{\bm a})\cdot {\bm \sigma}
\end{equation}
will denote the translated Dirac operator. The phase of the Dirac operator cannot be defined directly, because the latter has a zero eigenvalue so we need to use translates of $D$. The use of such translates is actually playing a crucial role for the analysis. So let ${\bm x}_0$ be a fixed point in $\mathbb{R}^d$. If ${\bm x}_0\notin \mathcal L$, we define the phase as:
\begin{equation}\label{F1}
F_{{\bm x}_0}=\frac{D_{{\bm x}_0}}{|D_{{\bm x}_0}|},
\end{equation}
which acts on $\mathcal{H}$ by multiplication with $\widehat{{\bm x}+{\bm x}_0}\cdot {\bm \sigma}$.  If ${\bm x}_0\in \mathcal L$, we define the phase as:
\begin{equation}\label{F2}
(F_{{\bm x}_0} \bm \psi)(\bm x)=\left \{
\begin{array}{l}
(\widehat{{\bm x}+{\bm x}_0}\cdot {\bm \sigma}) \bm \psi(\bm x), \ \mbox{if} \ {\bm x} \neq -{\bm x}_0 \bigskip \\
 ( \frac{1}{\sqrt{d}}\sum_{i=1}^d\sigma_i ) \bm \psi(\bm x) \ \mbox{if} \ {\bm x} = -{\bm x}_0.
\end{array}
\right .
\end{equation}
Clearly, for all cases, $F_{{\bm x}_0}$ has the required properties: 
\begin{equation}\label{DId}
(F_{{\bm x}_0})^\dagger = F_{{\bm x}_0}, \ (F_{{\bm x}_0})^2=1.
\end{equation}
In addition, one can show that, for any $f\in \mathcal A$, the operator $[F_{{\bm x}_0},\pi_\omega(f)]$ is compact. Summing up all the facts, we have demonstrated:

\begin{proposition}\label{OddFredholmModules} The triples $({\mathcal H},F_{{\bm x}_0},\pi_\omega)$, with ${\bm x}_0 \in [0,1]^d$ and ${\omega \in \Omega}$, define a family of odd Fredholm-modules over ${\mathcal A}$. \end{proposition}

\subsection{The natural Fredholm-modules in even dimensions}

This chapter describes the natural Fredholm-module over the non-commutative Brillouin torus in even dimensions $d$ \cite{ProdanJPA2013hg}. Let $\mathrm{Cliff}(d)$ denote the $2^{\frac{d}{2}}$-dimensional Hilbert space for the irreducible representation of the even $C_{d,0}$. The Hilbert space for the module is defined in the same way:
\begin{equation}
{\mathcal H}=\ell^2(\mathcal L,\mathbb{C}^N)\otimes \mathrm{Cliff}(d),
\end{equation}
and the $C^*$-algebra ${\mathcal A}$ is represented on ${\mathcal H}$ by the same $\pi_\omega \otimes 1$. The grading operator is taken to be 
\begin{equation}
\gamma = 1\otimes \gamma_0.
\end{equation}
The operator $F$ is defined again as the phase of the Dirac operator:
\begin{equation}
D=\sum_{i=1}^{d} X^i \otimes \gamma_i.
\end{equation}
The shorthands ${\bm v} \cdot {\bm \gamma}=\sum_{i=1}^{d} v^i \gamma_i $ and 
\begin{equation}
D_{\bm a}=({\bm X}+{\bm a})\cdot {\bm \gamma}
\end{equation}
will be used again. To define $F$, one actually has to consider again translates of the Dirac operator. Let ${\bm x}_0$ be a fixed point in $\mathbb{R}^d$. If ${\bm x}_0\notin \mathcal L$, we define:
\begin{equation}\label{F1}
F_{{\bm x}_0}=\frac{D_{{\bm x}_0}}{|D_{{\bm x}_0}|},
\end{equation}
which acts on ${\cal H}$ by multiplication with $\widehat{{\bm x}+{\bm x}_0}\cdot {\bm \gamma}$.  If ${\bm x}_0\in \mathcal L$, we define:
\begin{equation}\label{F2}
(F_{{\bm x}_0}\psi)(\bm x)=\left \{
\begin{array}{l}
(\widehat{{\bm x}+{\bm x}_0}\cdot {\bm \gamma})\psi(\bm x), \ \mbox{if} \ {\bm x} \neq -{\bm x}_0 \medskip \\
 ( \frac{1}{\sqrt{d}}\sum_{i=1}^{d}\gamma_i ) \psi(\bm x) \ \mbox{if} \ {\bm x} = -{\bm x}_0.
\end{array}
\right .
\end{equation}
Clearly, for all cases, $F_{{\bm x}_0}$ has the following properties: 
\begin{equation}\label{DId}
(F_{{\bm x}_0})^2=1, \ F_{{\bm x}_0}\gamma=-\gamma F_{{\bm x}_0}.
\end{equation}
In addition, one can show that, for any $f\in \mathcal A$, the operator $i[\hat{D}_{{\bm x}_0},\pi_\omega(f)]$ is compact. Summing up all the facts, we have demonstrated:

\begin{proposition}\label{EvenFredholmModules} The quadruples, $({\mathcal H},F_{{\bm x}_0},\gamma,\pi_\omega)$, with $\bm x_0 \in [0,1]^d$ and $\omega \in \Omega$, define a family of even Fredholm-modules over ${\mathcal A}$. \end{proposition}

For $d=2$, this family of Fredholm-modules is identical to the one used in the work \cite{BELLISSARD:1994xj} on the Integer Quantum Hall Effect.

\section{The Chern characters and their pairing with the $K$-groups}

\subsection{The $d =$ odd case}

To define the Chern characters, one needs first to address the issue of summability and the first observation is that indeed the required summability condition happens over the sub-algebra of localized observables. 

\begin{proposition}[\cite{ProdanOddChernArxiv2014}] The family of odd Fredholm-modules $({\mathcal H},F_{{\bm x}_0},\pi_\omega)$ defined in Proposition~\ref{OddFredholmModules}, with ${\omega \in \Omega}$ and ${\bm x}_0 \in [0,1]^d$, is $(d+1)$-summable over the sub-algebra $\mathcal A_{\mathrm{loc}}$, i.e. for any $f \in \mathcal A_{\mathrm{loc}}$:
\begin{equation}\label{OddSummability}
\int_{[0,1]^d} d\bm x_0 \int_\Omega d\omega \ \mathrm{Tr}\{|[\pi_\omega(f),F]|^{d+1}\} < \infty.
\end{equation}
\end{proposition}

Comparing with the previous exposition of the Fredholm-modules, one will notice that the summability condition was slightly modified in order to accommodate a family of Fredholm-modules (as oppose to an individual module). The summability property expressed above is important because it enables one to introduce the Chern character for the {\bf entire family of odd Fredholm modules}.

\begin{definition} The cohomology class (in the cyclic cohomology of $\mathcal A_{\mathrm{loc}}$) of the following $(d+1)$-cyclic cocycle:
\begin{align}\label{NewOddChernCharacter}
& \ \ \ \ \ \  \widetilde \tau_d (f_0,f_1,\ldots,f_d) = \frac{i^{d+1}}{2^d} \times \\ 
&  \int\limits_{[0,1]^d} d \bm x_0 \int\limits_\Omega d\bm P (\omega) \ \mathrm{Tr}'\{\pi_\omega(f_0)[F_{\bm x_0},\pi_\omega(f_1)] \ldots [F_{\bm x_0},\pi_\omega(f_d)]\}\nonumber
\end{align}
is called the odd Chern character of the family of odd Fredholm-modules $(\mathcal H, F_{\bm x_0},\pi_\omega)$. It will be denoted by $\widetilde{\mathrm{Ch}}_*$ in the following.
\end{definition}

\begin{theorem}[\cite{ProdanOddChernArxiv2014}]
The newly defined odd Chern character has the following fundamental properties:
\begin{enumerate}
\item The pairing between $\widetilde{\mathrm{Ch}}_*$ and $K_1^\mathrm{A}(\mathcal A_{\mathrm{loc}})$ remains integral:
\begin{equation}
\langle [v],\widetilde{\mathrm{Ch}}_* \rangle = \mathrm{Index} \ E_{\bm x_0} \pi_\omega (v) E_{\bm x_0} \ \in \mathbb{Z},
\end{equation}
where the Fredholm index on the right is almost surely independent of $\omega$ and $\bm x_0$.
\item The odd Chern cocycle accepts the following local formula:
\begin{align}
\widetilde \tau_d(f_0,f_1,\ldots,f_d)  =   \widetilde \Lambda_d \sum_{\rho \in S_d}(-1)^\rho {\mathcal T} \left ( f_0\prod_{i=1}^d \partial_{\rho_i}f_i\right ),
\end{align}
with $\widetilde \Lambda_d=\frac{i(-i\pi)^\frac{d-1}{2}}{d!!}$. $S_d$ is the permutations group.
\item The odd Chern cocycle $\widetilde \tau_d$ is continuous on $\mathcal A_{\mathrm{loc}}$. As such, the topological $K_1^\mathrm{T} (\mathcal A_{\mathrm{loc}})$ group can be used instead of the algebraic $K_{1}^\mathrm{A} (\mathcal A_{\mathrm{loc}})$ group.
\end{enumerate}
\end{theorem}
\proof Here are the key points of the proof. 

1. From the general theory presented in the first part of our exposition, one knows that for a fixed $\omega$ and a fixed invertible $v$, the numerical invariant resulted from the pairing of $\widetilde{\mathrm{Ch}}_*(\mathcal H, F_{\bm x_0},\pi_\omega)$ with $[v]$ is integral (with probability 1 in $\omega$ and $\bm x_0$). What we need to show is that this pairing remains integral after we average over $\omega$ and $\bm x_0$. This is facilitated by the fact that pairing of $\widetilde{\mathrm{Ch}}_*(\mathcal H, F_{\bm x_0},\pi_\omega)$ with the $K_1$ group is given by a Fredholm index (see Eq.~\ref{OddIntegrality}), and the latter is invariant under compact perturbations. Then one observes that, by substituting $\pi_\omega$ by a translate $\pi_{\mathfrak{t}_a \omega}$ in Eq.~\ref{OddIntegrality}, only a compact perturbation is produced on the righthand side of the equation, hence the paring of $\widetilde{\mathrm{Ch}}_*(\mathcal H, F_{\bm x_0},\pi_\omega)$ with the $K_1$ group remains unchanged. But the translations act ergodically on $\Omega$, hence this invariance to translations of $\omega$ implies that the pairing is constant of $\omega$, with the exception of singular cases which occur with zero probability. Similar for the invariance with respect to $\bm x_0$. The conclusion is that the pairing remains integral even after one averages over $\omega$ and $\bm x_0$.

2. Expending the trace in Eq.~\ref{NewOddChernCharacter}, one obtains:
\begin{align}
\widetilde \tau_d & (f_0,f_1,\ldots,f_d) =\\
& \frac{i^{d+1}}{2^d} \sum\limits_{{\bm x}_i \in \mathcal L} \int_{\mathbb{R}^d} d{\bm x} \ \mathrm{tr}_\sigma \left \{\prod_{i=1}^d (\widehat{{\bm x}_i + {\bm x}} - \widehat{{\bm x}_{i+1}+{\bm x}} ) \cdot {\bm \sigma} \right \} \nonumber \\
 &\times \int_\Omega d \bm P (\omega) \ \mathrm{tr}_\alpha\left \{\pi_\omega(f_0 )\prod_{i=1}^d\chi_{{\bm x}_i}\pi_\omega(f_i)\chi_{{\bm x}_{i+1}} \right \},\nonumber
\end{align}
where $\chi_x = \sum_\alpha (e_{\bm x}^\alpha)^\dagger e_{\bm x}^\alpha$ is the projector onto the orbitals at site $\bm x$, and $\mathrm{tr}_\alpha$ and $\mathrm{tr}_\sigma$ are the traces over the orbitals at $\bm x = \bm 0$ and over the odd $\mathrm{Cliff}(d)$, respectively. In the first line, one can use the following remarkable identity \cite{ProdanOddChernArxiv2014}:
\begin{align}\label{KIdentity1}
\int_{\mathbb{R}^d} d{\bm x} \ \mathrm{tr}_\sigma \left \{\prod_{i=1}^d (\widehat{{\bm x}_i + {\bm x}} - \widehat{{\bm x}_{i+1}+{\bm x}} ) \cdot {\bm \sigma} \right \} \nonumber \\
= \frac{2^\frac{d+1}{2} (2\pi)^\frac{d-1}{2}}{\imath^\frac{d-1}{2}d !!} \sum_{\rho \in S_d} (-1)^\rho \prod_{i=1}^d x_i^{\rho_i}, 
\end{align}
to reduce the expression of $\widetilde \tau_d$ to
\begin{align}
\widetilde \Lambda_d \int\limits_\Omega d\bm P (\omega) \sum_{\rho \in S_d} (-1)^\rho 
 \mathrm{tr}_\alpha \left \{ \pi_\omega(f_0) \prod_{i=1}^d \imath [X_i^{\rho_i},\pi_\omega(f_i)]  \right \}, \nonumber
\end{align}
which is exactly the expression given in the statement of theorem, if one uses the rules of calculus from Eqs.~\ref{Rule1} and \ref{Rule2}.

3. Using the local formula, the continuity of the Chern cocycle follows from Proposition~\ref{TildeA'}.
\qed
 
\medskip One can now return to the general theory presented in the first part of the exposition and harvest all its fruits. In particular, the pairing of  $\widetilde{\mathrm{Ch}}_*$ with the $K_1^\mathrm{T} (\mathcal A_{\mathrm{loc}})$ group (see Eq.~\ref{Pairing1}), together with the local formula given above, define a numerical invariant on the homotopy classes of $K_1^\mathrm{T} (\mathcal A_{\mathrm{loc}})$ group, which can be rightfully called the odd non-commutative Chern number:
\begin{equation}\label{NCOddChernNr}
\widetilde{\mathrm{Ch}}_d(v) \stackrel{\text{\tiny def}}{=} \widetilde \Lambda_d  \sum_{\rho \in S_d} (-1)^\rho \ \mathcal{T} \left \{ \prod_{i=1}^d v^{-1} \partial_{\rho_i}v  \right \}.
\end{equation}
This is an integer which cannot be changed under continuous deformations of $v$ inside $\mathcal A_{\mathrm{loc}}$ (which is considered with the GNS norm).

The last remark here is that $\widetilde{\mathrm{Ch}}_d(v)$ can be expressed in the operator representation, where it takes the form ($V_\omega = \pi_\omega v$):
\begin{align}
\widetilde{\mathrm{Ch}}_d(v) = & \widetilde \Lambda_d \int\limits_\Omega d\bm P (\omega) \sum_{\rho \in S_d} (-1)^\rho \nonumber \\
& \times \mathrm{tr}_\alpha \left \{\prod_{i=1}^d V_\omega^{-1} \imath [X_{\rho_i},V_\omega]  \right \},
\end{align}
which in the absence of disorder and magnetic fields is just the real-space representation of the classic odd Chern number (see Eq.~\ref{OddChernNr1}) over the Brillouin torus \cite{MondragonShemArxiv2013ew}. 
 
 \subsection{The $d =$ even case}
 
 As in the odd case, we address first the issue of summability.
 
 \begin{proposition} The family of even Fredholm-modules $({\mathcal H},F_{{\bm x}_0},\gamma,\pi_\omega)$ defined in Proposition~\ref{EvenFredholmModules}, with ${\omega \in \Omega}$ and ${\bm x}_0 \in [0,1]^d$, is $(d+1)$-summable over the sub-algebra $\mathcal A_{\mathrm{loc}}$:
\begin{equation}\label{EvenSummability}
\int_{[0,1]^d} d\bm x_0 \int_\Omega d\omega \ \mathrm{Tr}\{|[\pi_\omega(f),F]|^{d+1}\} < \infty,
\end{equation}
for any $f \in \mathcal A_{\mathrm{loc}}$.
\end{proposition}

There is an important observation to be made here. In Ref.~\cite{ProdanJPA2013hg}, the maximal non-commutative space (a certain noncommutative Sobolev space) where the almost sure $(d+1)$-summability of $({\mathcal H},F_{{\bm x}_0},\gamma,\pi_\omega)$ modules takes place was determined using the Dixmier trace. However, that technique, which in many regards can be considered optimal, cannot be used to prove the summability of the entire family of Fredholm modules, as defined above. Nevertheless, the latter can be established on $\mathcal A_{\mathrm{loc}}$ by a direct evaluation of Eq.~\ref{EvenSummability}, as it was done for the odd case in Ref.~\cite{ProdanOddChernArxiv2014}. As for the odd case, the summability property of Eq.~\ref{EvenSummability} enables a definition of the Chern character for the {\bf entire family of even Fredholm-modules}:

\begin{definition} The cohomology class (in the cyclic cohomology of $\mathcal A_{\mathrm{loc}}$) of the following $(d+1)$-cyclic cocycle:
\begin{align}\label{NewEvenChernCharacter}
& \ \ \ \ \ \  \tau_d (f_0,f_1,\ldots,f_d) = \\ 
& \int\limits_{[0,1]^d} d \bm x_0 \int\limits_\Omega d\omega \ \mathrm{Tr}'\{\pi_\omega(f_0)[F_{\bm x_0},\pi_\omega(f_1)] \ldots [F_{\bm x_0},\pi_\omega(f_d)]\}\nonumber
\end{align}
is called the even Chern character of the family of even Fredholm-modules $(\mathcal H, F_{\bm x_0},\gamma,\pi_\omega)$. It will be denoted by $\mathrm{Ch}_*$ in the following.
\end{definition}

One should note that, although the expressions of the odd and even Chern cocycles are formally the same, the definition of $\mathrm{Tr}'$ is different for these two cases (cf. to the presentation in the first part).

\begin{theorem}
The newly defined even Chern character has the following fundamental properties:
\begin{enumerate}
\item The pairing between $\mathrm{Ch}_*$ and $K_0^\mathrm{A}(\mathcal A_{\mathrm{loc}})$ remains integral:
\begin{equation}
\langle [p],\mathrm{Ch}_* \rangle = \mathrm{Index} \ \pi^-_\omega(p) F_{\bm x_0} \pi^+_\omega(p) \ \in \mathbb{Z},
\end{equation} 
where the Fredholm index on the right is almost surely independent of $\omega$ and $\bm x_0$.
\item The even Chern cocycle accepts the following local formula:
\begin{align}
\tau_d(f_0,f_1,\ldots,f_d)  =   \Lambda_d \sum_{\rho \in S_d}(-1)^\rho {\mathcal T} \left ( f_0\prod_{i=1}^d \partial_{\rho_i}f_i\right ),
\end{align}
with $\Lambda_d=\frac{(2\pi \imath)^\frac{d}{2}}{(d/2)!}$.
\item The even Chern cocycle $\tau_d$ is continuous on $\mathcal A_{\mathrm{loc}}$. As such, the topological $K_0^\mathrm{T} (\mathcal A_{\mathrm{loc}})$ group can be used instead of the algebraic $K_0^\mathrm{A} (\mathcal A_{\mathrm{loc}})$ group.
\end{enumerate}
\end{theorem}
\proof The above is a slightly different reformulation of the results of Ref.~\cite{ProdanJPA2013hg}. Here are the key points of the proof. 

1. Follows from the same reasons as in the odd case.

2. Expending the trace in Eq.~\ref{NewEvenChernCharacter}, one obtains:
\begin{align}
\tau_d & (f_0,f_1,\ldots,f_d) =\\
& - \sum\limits_{{\bm x}_i \in \mathcal L} \int_{\mathbb{R}^d} d{\bm x} \ \mathrm{tr}_\gamma \left \{ \gamma_0 \prod_{i=1}^d (\widehat{{\bm x}_i + {\bm x}} - \widehat{{\bm x}_{i+1}+{\bm x}} ) \cdot {\bm \gamma} \right \} \nonumber \\
 &\times \int_\Omega d \bm P (\omega) \ \mathrm{tr}_\alpha\left \{\pi_\omega(f_0 )\prod_{i=1}^d\chi_{{\bm x}_i}\pi_\omega(f_i)\chi_{{\bm x}_{i+1}} \right \},\nonumber
\end{align}
with the same notation as before. There is again a remarkable geometrical identity \cite{ProdanJPA2013hg}:
\begin{align}\label{KIdentity2}
\int_{\mathbb{R}^d} d{\bm x} \ \mathrm{tr}_\gamma \left \{ \gamma_0 \prod_{i=1}^d (\widehat{{\bm x}_i + {\bm x}} - \widehat{{\bm x}_{i+1}+{\bm x}} ) \cdot {\bm \gamma} \right \} \nonumber \\
= - \frac{(2 \pi)^\frac{d}{2}}{\imath^\frac{d}{2} (d/2)!} \sum_{\rho \in S_d} (-1)^\rho \prod_{i=1}^d x_i^{\rho_i}, 
\end{align}
which can be used to reduce the expression of $\tau_d$ to
\begin{align}
\Lambda_d \int\limits_\Omega d\bm P (\omega) \sum_{\rho \in S_d} (-1)^\rho 
 \mathrm{tr}_\alpha \left \{ \pi_\omega(f_0) \prod_{i=1}^d \imath [X_i^{\rho_i},\pi_\omega(f_i)]  \right \}, \nonumber
\end{align}
which is exactly the expression given in the statement of theorem, if one uses the rules of calculus from Eqs.~\ref{Rule1} and \ref{Rule2}.

3. Using the local formula, the continuity of the even Chern cocycle follows from Proposition~\ref{TildeA'}.
\qed

\medskip One can now return again to the general theory presented in the first part of the exposition and draw some very general conclusions. The pairing of  $\mathrm{Ch}_*$ with the $K_0^\mathrm{T} (\mathcal A_{\mathrm{loc}})$ group (see Eq.~\ref{Pairing2}), together with the local formula given above, define a numerical invariant on the homotopy classes of $K_0^\mathrm{T} (\mathcal A_{\mathrm{loc}})$ group, which can be rightfully called the even non-commutative Chern number:
\begin{equation}\label{NCEvenChernNr}
\mathrm{Ch}_d(p) \stackrel{\text{\tiny def}}{=} \Lambda_d  \sum_{\rho \in S_d} (-1)^\rho \ \mathcal{T} \left \{ p\prod_{i=1}^d  \partial_{\rho_i}p  \right \}.
\end{equation}
This is an integer which cannot be changed under continuous deformations of $p$ inside $\mathcal A_{\mathrm{loc}}$ (which is again considered with the GNS norm).

The last remark here is that $\mathrm{Ch}_d(p)$ can be expressed in the operator representation, where it takes the form ($P_\omega = \pi_\omega P$):
\begin{align}
\mathrm{Ch}_d(p) = & \Lambda_d \int\limits_\Omega d\bm P (\omega) \sum_{\rho \in S_d} (-1)^\rho \nonumber \\
& \times \mathrm{tr}_\alpha \left \{P_\omega \prod_{i=1}^d  \imath [X_{\rho_i},P_\omega]  \right \},
\end{align}
which in the absence of disorder and magnetic fields is just the real-space representation of the classic even Chern number (see Eq.~\ref{EvenChernNr1}) over the Brillouin torus \cite{ProdanJPA2013hg}.
 
 \subsection{Discussion} 
 
 The quantization and Invariance of the odd/even Chern numbers can be proven by starting from the Fredholm indices
 \begin{equation}
 \mathrm{Index} \ E_{\bm x_0} V_\omega E_{\bm x_0} \ \mathrm{and} \ \mathrm{Index} \ P^-_\omega F_{\bm x_0} P^+_\omega,
 \end{equation}
 respectively. Indeed, the index of these operators can be evaluated using Fedosov's formula and the key identities of Eqs.~\ref{KIdentity1} and \ref{KIdentity2}. But without the guidance of non-commutative geometry, who would have guessed that these are the correct operators to start with? 
 
 If the author is allowed to share some thoughts about his experience, then these will be his words for the non-commutative geometry as applied to materials science:
 \begin{itemize}
 \item The framework provides guidance and intuition. When done inside this framework, the search for the correct invariants no longer feel like searching for a needle in a haystack. 
 \item It provides the big picture so one can always know what he is computing. In the present context, the index theorems we just presented give morphisms from the $K$-groups of algebra of localized observables into the $\mathbb Z$.
 \item Last but not the least, the framework provides some outstanding tools of calculus. It will be a true asset to the materials science if a wider acceptance is achieved among the physicists. Needles to say, the field of Topological Insulators is the perfect ground for applications.
 \end{itemize}
 
 \section{Stability of the complex topological phases}
 
 With the theory of the odd and even Chern numbers in place, the characterization of the topological phases of the A and AIII-symmetry classes of condensed matter systems, \emph{in the presence of strong disorder and magnetic fields}, can be fully accomplished.
 
 \subsection{The AIII-symmetry class}
 A homogeneous aperiodic system is in AIII-symmetry class if there exists a Hermitean $N \times N$ matrix $S$ acting on the orbital degrees of freedom, such that $S^2 =1$ and:
 \begin{equation}
 (1\otimes S) H_\omega (1\otimes S^{-1}) = - H_\omega \ \mbox{for all} \ \omega \in \Omega.
 \end{equation}
 
 It is always useful to summarize some fundamental properties:
 \begin{itemize} 
 \item The above symmetry constraint forces the energy spectrum to be symmetric relative to the origin. For the AIII-symmetry class, the Fermi level is always considered to at zero.
 \item Let
 \begin{equation}
S_\pm = \frac{1}{2}(1\pm S)
\end{equation}
denote the spectral projections of $S$ onto its $\pm 1$ eigenvalues. Then:
\begin{align}\label{Deco}
\Phi(H_\omega) = & (1\otimes S_-) \Phi(H_\omega)  (1\otimes S_+) \\
& + (1\otimes S_+) \Phi(H_\omega) (1\otimes S_-), \nonumber
\end{align}
for any odd function $\Phi$. 
\item The ranks of $S_\pm$ are equal. As such, there exists a unitary matrix $R$ such that:
\begin{equation}
R S_\pm R^{-1} = S_\mp.
\end{equation}
\end{itemize}
 
As detailed in Ref.~\cite{ProdanOddChernArxiv2014}, the ground states of these Hamiltonians can be uniquely characterized by a family of unitary operators $\{U_\omega\}_{\omega \in \Omega}$ on $\ell^2(\mathcal L)\otimes \mathbb C^N$. Indeed, the ground state of $H_\omega$ is in one to one relation to the phase of the Hamiltonian:
\begin{equation}
Q_\omega = \frac{H_\omega}{|H_\omega|} = \mathrm{sign}(H_\omega),
\end{equation}
because the projector onto the occupied electron states can be obtained as $\frac{1}{2}(1-Q_\omega)$. For $Q_\omega$ to be well defined, one must assume that, with probability one in $\omega$, the origin is not an eigenvalue of $H_\omega$, which is the case if the energy spectrum is void or localized at the Fermi level.

Now, using $Q_\omega^2 =1$, Eq.~\ref{Deco} and the definition of the $R$ matrix, one can immediately see that:
\begin{align}\label{UOmega}
U_\omega = [(1\otimes R) Q_\omega (1\otimes S_-)] \oplus (1\otimes S_+) 
\end{align}
is indeed a unitary matrix on $\ell^2(\mathcal L)\otimes \mathbb C^N$.

Another good news is that all the important objects can be ported on the non-commutative Brillouin torus. Indeed, $1\otimes S$ is generated by $s(\omega,\bm x) = \delta_{\bm x, \bm 0}S$, $1\otimes S_\pm$ by $s_\pm(\omega,\bm x) = \delta_{\bm x, \bm 0}S_\pm$ and $1\otimes R$ by $r(\omega,\bm x) = \delta_{\bm x, \bm 0}R$. Then $U_\omega = \pi_\omega (u)$ where:
\begin{equation}
u = \big(r \ \mathrm{sign}(h) \ s_-\big)+s_+.
\end{equation}
As such, we have established that the ground state of the homogenous aperiodic condensed matter systems from the AIII-symmetry is in one to one correspondence with an unitary element, which is in the algebra ${\mathcal A}$ if a spectral gap exists at the Fermi level, and is in the weak von Neumann closure ${\mathcal A}'$ if the this gap is close but the spectrum is localized at the Fermi level. In this later case, one can actually prove much more \cite{ProdanOddChernArxiv2014}, that the kernels of $U_\omega$'s, when averaged over disorder, decay exponentially fast. In other words, $u$ belongs to the sub-algebra $\mathcal A_{\mathrm{loc}}$ of the localized observables and, as such, its odd Chern number is well defined. Furthermore, under continuous deformations of the Hamiltonians, the element $u$ varies continuously with respect to the GNS-norm. As such, all the conclusions from the previous chapter apply in their entirety and one can immediately conclude:

\begin{corollary}[Stability of the Topological Phases in the AIII-symmetry class \cite{ProdanOddChernArxiv2014}]  
Consider a homogeneous topological insulator from the AIII-symmetry class. Then:
\begin{enumerate}
\item If the Fermi level resides in a region of localized energy spectrum, the non-commutative odd-Chern number corresponding to its ground state is finite and quantized:
\begin{equation}
\widetilde{\mathrm{Ch}}_d(u) = \mathrm{Index}\ E_{{\bm x}_0}U_\omega E_{{\bm x}_0} \in \mathbb Z.
\end{equation}
\item Let $H_\omega(t)=H_\omega+t \delta H_\omega$, with $\sup |(e_{\bm x}^\alpha |\delta H_\omega|e_{\bm y}^\beta)|\leq \infty$, be a deformation of the homogeneous system preserving the AIII-symmetry. Assume that the Fermi level stays in a region of localized spectrum at all times. Then the odd-Chern number corresponding to the ground state of $H_\omega(t)$, that is $\widetilde{\mathrm{Ch}}_d(u_t)$, remains pinned at a quantized value for all $t$'s.   
\end{enumerate}
\end{corollary}

\subsection{The A-symmetry class}

The characterization of the topological phases from the A-symmetry class is much simpler because, in the absence of any symmetry constraints (other than the self-adjointness of $H_\omega$), the ground state of a homogeneous system is encoded directly in the family of projectors $P_\omega$ onto the energy spectrum below $E_F$. This projectors are generated by the element $p=\frac{1}{2}(1-\mathrm{sign}(h))$ and, given the discussion from the previous section, it follows that, under Anderson localization condition, $p$ belongs to the sub-algebra $\mathcal A_{\mathrm{loc}}$ of localized observables. Furthermore, under continuous deformations of the Hamiltonians, $p$ varies continuously with respect to the GNS-norm. As such, the conclusions from the previous chapter apply again in their entirety and one can conclude:

\begin{corollary}[Stability of the Topological Phases in the A-symmetry class \cite{ProdanJPA2013hg}]Consider a homogeneous topological insulator from the A-symmetry class. Then:
\begin{enumerate}
\item If the Fermi level resides in a region of localized energy spectrum, the non-commutative even-Chern number corresponding to its ground state is finite and quantized:
\begin{equation}
\mathrm{Ch}_d(p) = \mathrm{Index}\ P^-_\omega F_{{\bm x}_0}P^+_\omega \in \mathbb Z.
\end{equation}
\item Let $H_\omega(t)=H_\omega+t \delta H_\omega$, with $\sup |(e_{\bm x}^\alpha |\delta H_\omega|e_{\bm y}^\beta)|\leq \infty$, be a deformation of the homogeneous system. Assume that the Fermi level stays in a region of localized spectrum at all times. Then the even-Chern number corresponding to the ground state of $H_\omega(t)$, that is $\mathrm{Ch}_d(p_t)$, remains pinned at a quantized value for all $t$'s.   
\end{enumerate}
\end{corollary}

\section{Discussion}

Let us first summarize what the above mathematics has given us. First, we now can be absolutely sure that \emph{all} the topological phases of the condensed matter systems from the A- and AIII-symmetry classes, appearing in the presently accepted classification table of topological insulators \cite{SchnyderPRB2008qy,kitaev:22,RyuNJP2010tq}, continue to exist even after the insulating gaps are closed by disorder and filled with dense localized spectrum. Furthermore, we can also be absolutely sure that, at the crossing from one topological phase to another, an Anderson localization-delocalization transition takes place. For the experiment, this means that the phase boundaries of different topological phases are sharp and can be mapped using transport experiments. The experimental signature of the phase boundary should be exactly the same as the one recorded at the plateau-plateau transitions in the Integer Quantum Hall Effect (we refer to the spike in the direct conductance and its universal scaling with temperature). On the practical side, the non-commutative formulas for the invariants can be implemented on a computer \cite{ProdanAMRX2013bn}, leading to some of the most accurate and efficient computational algorithms for phase diagram mappings \cite{ProdanJPhysA2011xk,MondragonShemArxiv2013ew,SongArxiv2014bb}.

The topological invariants labeling different phases do not always have direct physical interpretations in terms of measurable bulk quantities. For the A-symmetry class, the even Chern numbers can be connected to electromagnetic response coefficients, such as the linear Hall conductance in $d=2$, or to the non-linear (second-order) electromagnetic response coefficient defined in \cite{ZhangScience2001jv} for $d=4$. The odd Chern number (taken mod 2) gives the electric polarization in $d=1$ \cite{MondragonShemArxiv2013ew} but a direct physical interpretation of the invariant (i.e. not mod 2) is probably impossible in $d=1$ and beyond. Also, it is generally accepted that the bulk $\bm Z_2$ invariant for the time-reversal symmetric topological insulators, which are presently scrutinized in many laboratories, has no direct physical interpretation. As a general rule, however, the presence of a nontrivial bulk invariant will have measurable effects for the physics at the boundary between a bulk sample and the vacuum \cite{Qi:2008cg}. This bulk-boundary correspondence is a separate problem and we are not ready yet to make any statements about it. 

One important remaining task for the bulk characterization of the complex classes to topological insulators is the computation of the $K$-groups of the local algebra $\mathcal A_{\mathrm{loc}}$. It remains to be seen if these $K$-groups are isomorphic with $\mathbb Z$, in which case the non-commutative Chern numbers will give a full classification of the complex classes.

\paragraph{Acknowledgments} 
I would like to thank my collaborators who made possible this journey in non-commutative geometry. They are Bryan Leung, Hermann Schulz-Baldes and Jean Bellissard. This work was supported by the U.S. NSF grants DMS-1066045 and DMR-1056168.

\bibliography{../../../TopologicalInsulators}
\end{document}